\documentclass{jpsj-suppl}
\usepackage{txfonts, xspace} 
\newcommand\nuenueb{$\nu_e / \bar{\nu}_e$\xspace}
\newcommand\numunumub{$\nu_\mu / \bar{\nu}_\mu$\xspace}
\newcommand\nue{$\nu_e$\xspace}
\newcommand\numu{$\nu_\mu$\xspace}
\newcommand\piz{$\pi^0$\xspace}

\newcommand\gammapuraftergammacuts{96.6\xspace}

\newcommand\oofvgammacuts{75.6\xspace}
\newcommand\CCcontaminationgammacuts{61.9\xspace}
\newcommand\puraftergammacuts{~0.01\xspace}
\newcommand\CCcontaminationCCcuts{54.2\xspace}
\newcommand\xfgd{186\xspace}
\newcommand\yfgd{186\xspace}
\newcommand\zfgd{33\xspace}
\newcommand\oofvaftervetocuts{46.7\xspace}
\newcommand\purafterallcuts{~0.03\xspace}

\title{A Search for Neutral Current Single Gamma with ND280 at T2K}

\author{Pierre \textsc{Lasorak}$^{1}$, for the T2K Collaboration}

\inst{$^{1}$ Particle Physics Research Center, School of Physics and Astronomy, Queen Mary University, London, E1 4NS, United Kingdom}

\email{p.j.j.lasorak@qmul.ac.uk}

\recdate{January 4, 2015}

\abst{The methodology and preliminary results for the search of single photons initiated by Neutral Current neutrino interactions with the ND280 detector at the T2K experiment
are presented.
This measurement aims to set the first limit on single-photon neutrino production below 1~GeV. Neutrino production of single photon is a subdominant process in neutrino interactions.
Because photons and electrons have very similar signatures in neutrino detectors, careful estimations need to be made not to bias the \nue appearance oscillation results of accelerator
neutrino experiments. The single photons are created by a nuclear resonance (typically $\Delta$(1232)) after interaction of the neutrino. The cross section is expected to
be of the order
of $10^{-42}$~cm$^{2}$. The main background is composed of \piz decaying into two photons, where only one photon is detected, and \piz events creating photons from outside of
the fiducial volume.}

\kword{NCgamma, neutrino, neutral current interaction, cross section, photon emission, resonance excitation, neutrino experiment}

\begin{document}
\maketitle

\section{Introduction}
\begin{figure}[tbh]
\begin{center}
\includegraphics[width=0.6\textwidth]{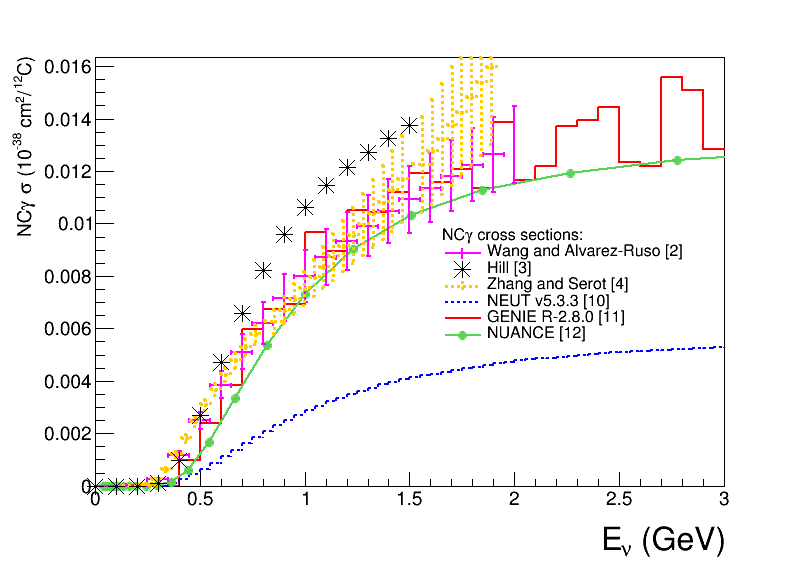}
\caption{NC gamma cross section from various models~\cite{LARCalculation,Hill,Serot} and generators~\cite{NEUT,GENIE,NUANCE} (Colour online).}
\label{fig:crosssections}
\end{center}
\end{figure}

In Cherenkov detectors, photons and electrons are complicated to distinguish, see, for example, Table~I of~\cite{MiniBooNElowE}.
It is therefore very important for long baseline experiment measuring electron (anti-) neutrino (\nuenueb) appearance in muon (anti-) neutrino (\numunumub)
beam to estimate the rate of the processes in which a \numu can create photons by Neutral Current Interaction (NC).
The object of this analysis is to search for processes where a single photon and no lepton / meson is emitted after neutrino NC interaction (except the neutrino),
later referred to as ``NC gamma''.
This is expected to be a rare process, cross section calculations predict $\simeq 10^{-42}$~cm$^{2} /$~nucleon~\cite{LARCalculation,Hill,Serot,Rosner},
they are shown in Figure~\ref{fig:crosssections}.

To achieve this, the Fine Grain Detector (FGD) of the off-axis near detector (ND280) of T2K~\cite{T2KExperiment} is used. It is located 280~m away from the target station at J-PARC and,
at this location, the muon neutrino beam is peaked at 600~MeV~\cite{T2KFlux}.

T2K has measured with unprecedented accuracy and significance the mixing parameters
$\theta_{23}$ and $\Delta m^2_{23}$~\cite{T2KLongOscillationPaper},
and produced the best measurement of $\theta_{23}$ using a neutrino beam and Super-Kamiokande (SK)~\cite{SKExperiment}.

This note starts with a description of the selection of single photon events in Section~\ref{seq:selection}.
The component of the selection are then explained in Section~\ref{seq:backgrounds}. The preliminary results
are shown in Section~\ref{seq:sensitivity}. Finally, the on-going work in neutrino generator is detailed in Section~\ref{seq:genie}.

\section{Selection of Events} \label{seq:selection}
The basic idea of the selection is to identify exclusive neutrino NC gamma neutrino interaction within the fiducial volume of the FGD.
The topology of the event corresponds to a photon converting on the Carbon of the FGD, creating a electron/positron pair that are propagating inside the TPC.
The selection first focuses on identifying photons and then veto the events in which there is another meson or charged lepton (such as \piz, $\pi^\pm$, $\mu^\pm$, ...).

\subsection{Photon selection}
The first requirement is to have two tracks in the FGD and in the TPC with opposite charge, and energy loss ($dE/dx$) consistent with the electrons.
The next cut is forcing the tracks to start at a distance closer than 10~cm from each other. 
The accurate 3-momentum reconstruction of the TPC is then used to reconstruct the invariant mass of the electron-positron system, which is required to be smaller than 50~MeV.

At this stage, the sample is composed at more than \gammapuraftergammacuts\% of photons converting in the FGD (Note that this are not necessary NC gamma events and that
all the numbers in this section are produced using predictions from NEUT v5.3.2~\cite{NEUT} and the official T2K Monte Carlo for ND280).
The majority of the sample is composed of interactions where the detected photon comes from one \piz.
The neutrino interaction occurred \oofvgammacuts\% of the time outside the fiducial volume of the FGD (OOFV, for Out Of Fiducial Volume).
More than a half of the interactions are coming from Charged Current (CC) interactions.
The single photon events (NC gamma) are representing \puraftergammacuts\% of the selection after these cuts.

\subsection{Single photon selection}
To single out NC gamma interactions, a first step is to remove the CC interactions. This is done by vetoing muons: the $dE/dx$ of all the tracks that are going through
the TPCs are reconstructed and those consistent with the muon hypothesis cause the event to be rejected.
This background is further reduced by searching for long tracks, which propagate in both the surrounding Electromagnetic Calorimeter (ECal) and the Side Muon Range Detector (SMRD).
After performing this cut, the CC background is decreased from \CCcontaminationgammacuts\% to \CCcontaminationCCcuts\%.

The other significant background in the photon selection described before is the OOFV events.
Given that the photon usually propagates 40~cm in the plastic and that the size of the FGD is
\xfgd~cm~$\times$~\yfgd~cm~$\times$~\zfgd~cm, this is not surprising. To get a good rejection of these events, one would need to have a large homogeneous detector and remove
$\simeq$~40~cm on each side to select only event occuring inside the detector.
In the case of the present analysis, an ``upstream veto'' is performed: any event which produces reconstructed object in the upstream \piz detector (P0D) are vetoed.
To ease the background selection (see in the next section), a second ``cone veto'' is realised by searching for object starting (or ending) in the surrounding ECal detector
in a cone in the backward direction of the reconstructed photon.
The OOFV contamination is diminished from \oofvgammacuts\% to \oofvaftervetocuts\%.

The last background comes from \piz neutrino-production inside and outside the FGD.
To reduce this background, the requirement is that there is no reconstructed object in the ECal downstream of the photon.
Using this, one should expect to detect the second photon from \piz decay. This also rejects multi-pions events when the detected photon 
comes from a \piz and a charged pion propagates in the ECal. 

\begin{figure}[tbh]
\begin{center}
\includegraphics[width=0.6\textwidth]{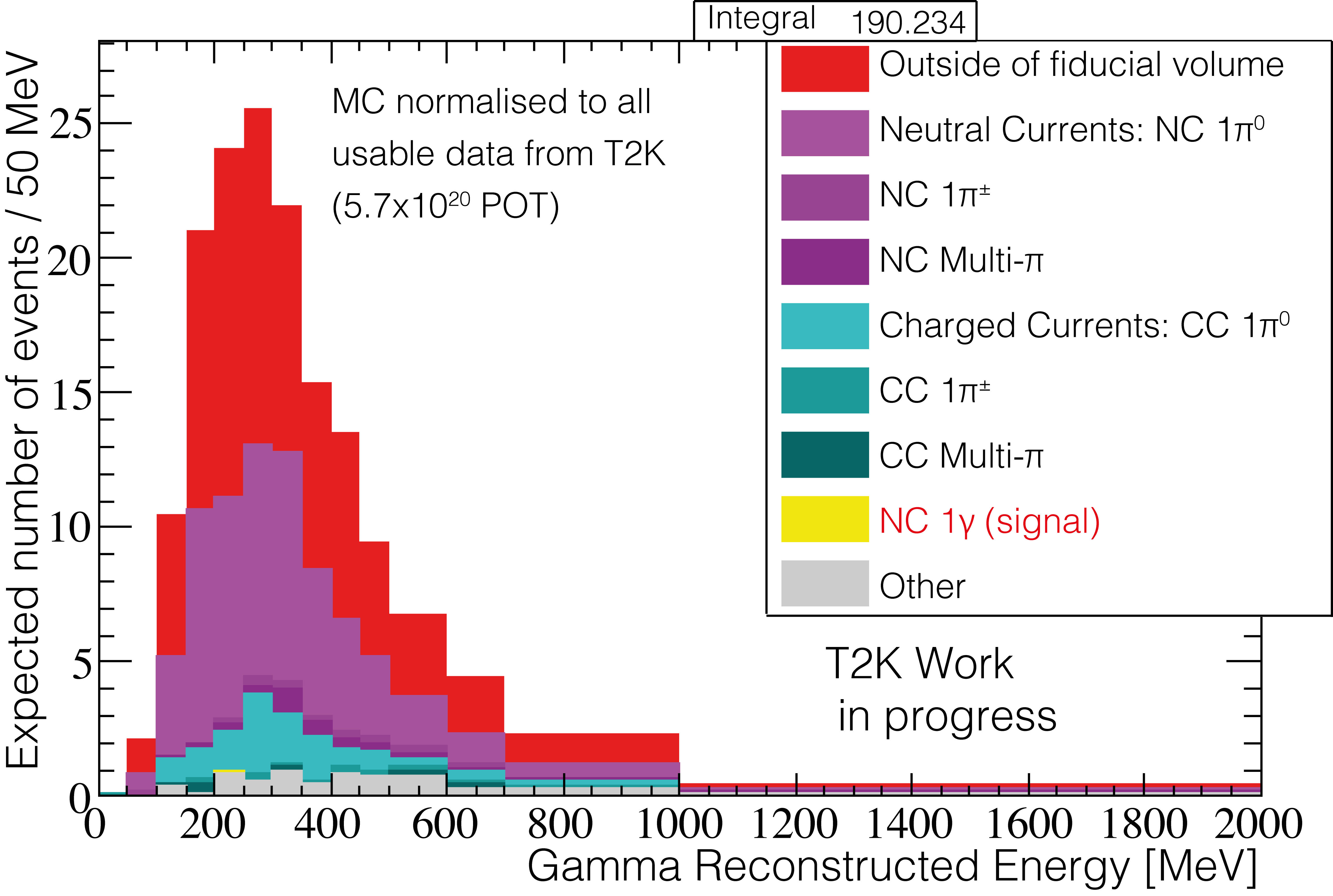}
\caption{Reconstructed energy of the selected photons. The selection is largely dominated by photon from \piz both coming from inside and outside the Fiducial Volume of the FGD (Colour online).}
\label{fig:energy}
\end{center}
\end{figure}

After these vetoes, the signal purity is \purafterallcuts\%, the reconstructed energy distributions of the signal is shown on Figure~\ref{fig:energy}.

\section{Sample description} \label{seq:backgrounds}
As expected by the smallness of the cross section, the selection is largely background-dominated. Therefore the limit is driven by the uncertainties on the background.
Careful characterisations of these are required to select appropriate control samples.

As stated before, the OOFV events are one of the dominant backgrounds. There are large areas of dead material around the FGD and an uncertainty of 40~cm
on the neutrino vertex leads to a large fraction of selected photon that are coming from these areas. Unfortunately, there are few analysis in the ND280 that
rely on the accurate modelling of these areas. One could therefore expect that mis-modellings would still be present in the geometry implemented for Monte Carlo generation. 
These errors could eventually lead to significant discrepancies in this sample.
It is also very hard to select this kind of interaction. The systematic uncertainty associated with
these events is still under development, however one can use the cone veto described before to get a sense of the agreement between the simulation and data.

The second background is the \piz events. After checking the kinematics of the \piz events remaining in the selection after all the cuts, it can be deduced
that they are primarily composed of asymmetric decay of \piz. In other words, the secondary photon has low energy. The ECal detection efficiency of
photons gets lower for energy smaller than 100~MeV, which makes these photons invisible to the ND280. The second topology that was observed in these \piz events is when the
secondary photon converts in dead material, which again complicates their selection.

The systematics on the backgrounds have been evaluated using approximations to get a sense of the sensitivity to this channel.
The largest systematics on the background were considered and added in quadrature.
Keeping in mind that these backgrounds are going to be constrained by the control samples for the full analysis, this is likely to be an overestimation. The main backgrounds
correspond to \piz production and OOFV contamination. In previous T2K analysis~\cite{NueMeasurementT2K}, the errors on these were both assigned to 30\%.

\section{Expected sensitivity} \label{seq:sensitivity}
To get the sensitivity, the ND280 simulations were used and scaled to the T2K Proton on Target (PoT) that one can use for this analysis, corresponding to good data quality
across all the ND280 and with all sub-detectors in place. From this, one can evaluate the number of expected events in the selection and therefore limit on
cross section for the channel.

\begin{figure}[tbh]
\begin{center}
\includegraphics[width=0.6\textwidth]{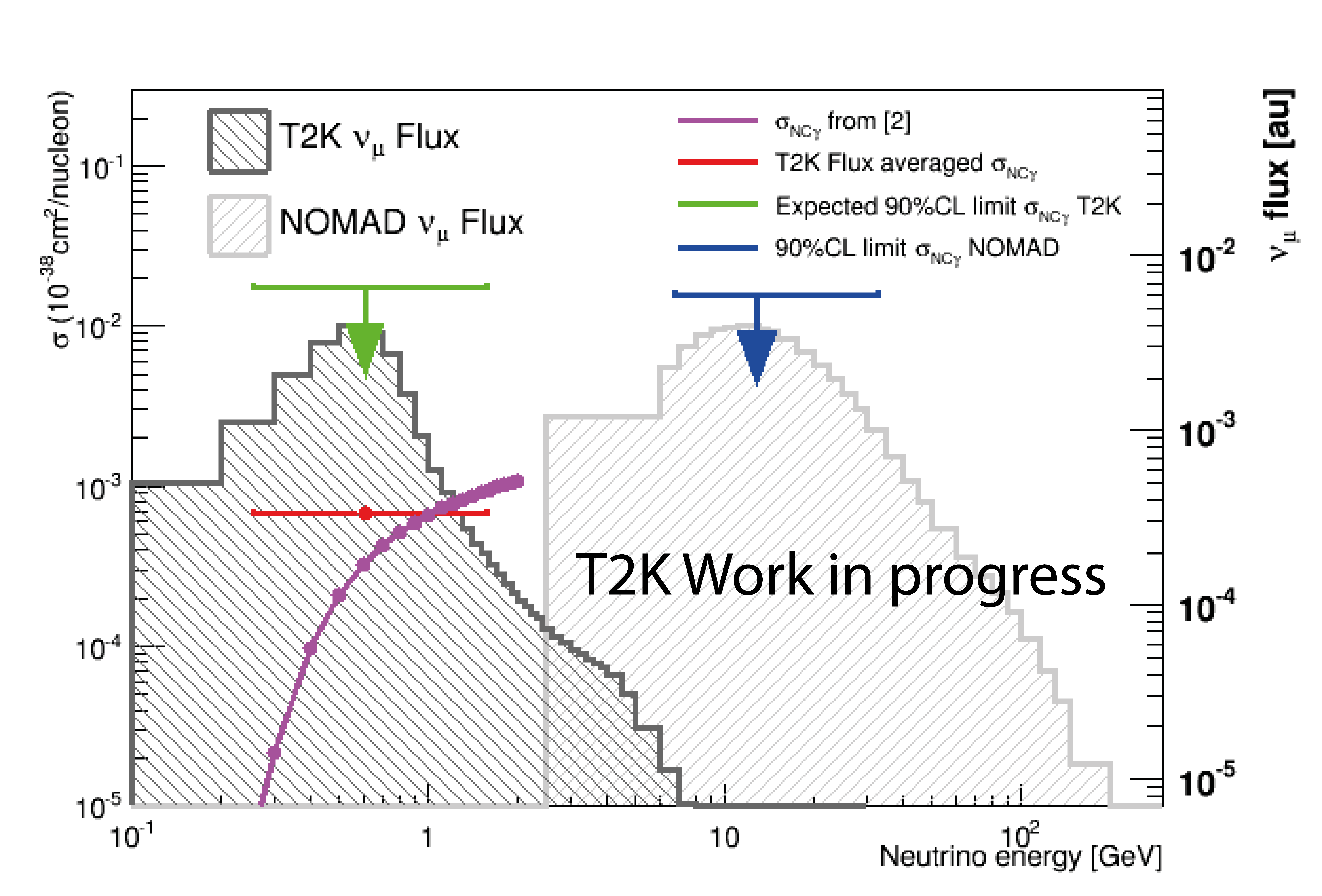}
\caption{ND280 sensitivity (Monte Carlo only) to NC$\gamma$ (green) along with the cross section from~\cite{LARCalculation} (purple and red).
The limit from NOMAD~\cite{NOMADNCg} is included (blue). T2K~\cite{T2KFlux} and NOMAD fluxes~\cite{NOMADFlux} are shown (grey).}
\label{fig:sensitivity}
\end{center}
\end{figure}

In Figure~\ref{fig:sensitivity}, the ND280 sensitivity is shown along with the published limit from NOMAD~\cite{NOMADNCg,NOMADFlux}.
The limit is one order of magnitude higher than the flux-integrated cross-section from~\cite{LARCalculation}, illustrating the difficulties to perform this measurement.

The cross section limit shown in Figure~\ref{fig:sensitivity} is extracted using the formula that was used for the extraction of the cross section of the homonym channel at SK~\cite{SKNCg}:
\begin{equation}\label{eq:xsec}
\langle \sigma^\text{obs}_{\nu, \text{NC}\gamma} \rangle_\text{90\%CL} = \frac{Q^\text{obs}_\text{90\%CL}-N_\text{bkg}^\text{exp}}{N^\text{exp}-N_\text{bkg}^\text{exp}}\langle \sigma^\text{NEUT}_{\nu, \text{NC}\gamma} \rangle,
\end{equation}
where $\langle \sigma^\text{obs}_{\nu, \text{NC}\gamma} \rangle_{90\%CL}$ is the 90\% confidence limit on the upper value of the cross section and $\langle \sigma^\text{NEUT}_{\nu, \text{NC}\gamma} \rangle$ is the flux-averaged cross section from NEUT~\cite{NEUT}. The total number of expected events is $N^\text{exp}$, and $N^\text{exp}_\text{bkg}$ are the expected number of background events, respectively 

The total number of observed events is $N^\text{obs}$ and is linked to the 90\% upper quantile $Q_\text{90\%}$ by the relation:

\begin{equation}\label{eq:quantile}
0.9 = \int_{-\infty}^{Q_\text{90\%}} \text{Gauss}\left(N^{\text{obs}}, \epsilon\right) dN
\end{equation}

Where $\epsilon$ is the systematic uncertainty of the selection in a single photon momentum bin.

To get the Figure~\ref{fig:sensitivity}, one needs to apply all the systematic
variations to the selection, which gives the spread in number of events of the selection. This distribution is expected to be Gaussian, and was treated as such for this letter,
but the full systematic uncertainties treatment will not rely on this assumption. Then, the average number of background events (non NC gamma) is substracted. This shifts the
distribution to be around the number of NC gamma events one expects in the selection. Since the error on this number is expected to be bigger than the number of NC gamma events,
the upper 90\% quantile of the distribution (from Equation~\ref{eq:quantile}) was estimated, and is used in Equation~\ref{eq:xsec}.

\section{NC gamma in GENIE} \label{seq:genie}
Parallel to searching for NC gamma with the ND280, there is an on-going effort to implement the model of~\cite{LARCalculation} in GENIE~\cite{GENIE}.
The improvements that this would lead to are highlighted here. Other models were developed in~\cite{Hill, Serot, Rosner}.

The neutrino generators, like NEUT and GENIE, usually rely on the assumption that the photon comes from the decay of nucleon resonances.
As for the pion production, the resonance $\Delta$(1232) is the most dominant in all models.
However for such exclusive channels, it is important to take into account the interferences between the resonances and also the ``background'' contributions (non-resonant
channels). At this point, GENIE does not include these: it is using the Rein and Sehgal model~\cite{ReinSehgal} and low energy DIS to take into account
the background contributions (which does not produce single photon). NEUT uses the Rein and Sehgal model with partial background contribution (isospin 1/2 only).

Another limit is that the generator does not include the RPA (Random Phase Approximation) screening effects for these processes, so different behaviour is expected at low $Q^2$.

Furthermore, in GENIE (and NEUT), the resonances decay electromagnetically with a constant branching ratio compared to the one of the pion decay.
However, this branching ratio is expected to depend on the invariant mass of the resonance.
Since the masses of the resonances obey Breit-Wigner distributions, a simple phase space argument shows that the fraction of decays to photon
is expected to increase compared to the one of the decays to pion (because the pion is massive).

Finally, while interacting, the neutrino creates a polarised Z boson, which in its turns polarises the resonance.
Therefore, the decay of the resonance is anisotropic, which is neglected in the case of generators.
However since the resonance is boosted, the difference on the actual final state topology is expected to be small for most of the cases. One
can compare, for example, the Figure~5 of~\cite{KendallLAR}, where one sees that the difference between NEUT and~\cite{LARCalculation} are overlaid in the photon angle space.

\end{document}